\documentclass[useAMS, usenatbib, usegraphicx]{mn2e}
\usepackage{times}
\usepackage{color}

\title[Star formation in the Scutum tangent region]{Star formation towards the Scutum tangent region and the effects of Galactic environment}
\author[D. J. Eden et al.]{D. J. Eden$^{1}$\thanks{E-mail: dje@astro.livjm.ac.uk}, T. J. T. Moore$^{1}$, R. Plume$^{2}$ and L.K. Morgan$^{1}$\\
$^{1}$Astrophysics Research Institute, Liverpool John Moores University, Twelve Quays House, Egerton Wharf, Birkenhead CH41 1LD\\
$^{2}$Department of Physics and Astronomy, University of Calgary, 2500 University Drive NW, Calgary, Alberta T2N 1N4, Canada}
\begin{document}

\date{Accepted. Received; in original form}

\pagerange{\pageref{firstpage}--\pageref{lastpage}} \pubyear{2012}

\maketitle

\label{firstpage}

\begin{abstract}

By positional matching to the catalogue of Galactic Ring Survey molecular clouds, we have derived distances to 793 Bolocam Galactic Plane Survey (BGPS) sources out of a possible 806 located within the region defined by Galactic longitudes $\emph{l}$ = 28.5$\degr$ to 31.5$\degr$ and latitudes $\mid$\emph{b}$\mid$ $\leq$ 1$\degr$. This section of the Galactic Plane contains several major features of Galactic structure at different distances, mainly mid-arm sections of the Perseus and Sagittarius spiral arms and the tangent of the Scutum-Centarus arm, which is coincident with the end of the Galactic Long Bar. By utilising the catalogued cloud distances plus new kinematic distance determinations, we are able to separate the dense BGPS clumps into these three main line-of-sight components to look for variations in star-formation properties that might be related to the different Galactic environments. We find no evidence of any difference in either the clump mass function or the average clump formation efficiency (CFE) between these components that might be attributed to environmental effects on scales comparable to Galactic-structure features.

Despite having a very high star-formation rate, and containing at least one cloud with a very high CFE, the star formation associated with the Scutum-Centarus tangent does not appear to be in any way abnormal or different to that in the other two spiral-arm sections. Large variations in the CFE are found on the scale of individual clouds, however, which may be due to local triggering agents as opposed to the large-scale Galactic structure.

\end{abstract}


\section{Introduction}

Observations have shown that the star-formation efficiency (SFE) on large scales is generally low ($\sim$ 1 per cent; \citealt*{b18}) in the Galaxy and other normal galaxies. However, in starburst galaxies the SFE can increase by $\sim$ 50 times \citep*{b19}. Triggering mechanisms are thought to be the cause of increases in the SFE on smaller scales in some Galactic star-forming regions (W3: \citealt{b20}). Such changes may depend on the turbulent Mach number within the cloud \citep{b22}, local triggering and feedback mechanisms, such as expanding H$\, \textsc{II}$ regions \citep*{b11}, or proximity to features of the large-scale structure of the Galaxy.

Star formation is observed to be associated with spiral arms and other features of Grand Design structures in our Galaxy and other spiral galaxies. It is possible that the spiral arms are assembling the star-forming regions so that feedback and triggering mechanisms become more important within them \citep{b12}. However, molecular cloud formation theories postulate that the spiral density waves are responsible for forming the clouds \citep{b13}. Observations of the outer-part of the Perseus spiral arm by \citet{b14} have shown that the fraction of interstellar gas in molecular form is much higher within the spiral arm than the inter-arm region. This supports the theory that the atomic gas is compressed upon entering the spiral shock, allowing molecular gas to condense and suggests that the spiral arms are triggering additional star formation through the formation of new clouds. A more recent study by \citet{b6}, using the Galactic Ring Survey (GRS), found similar results, implying that the clouds in the inter-arm regions dissipate quickly and that the spiral arms are indeed responsible for forming the clouds.

The spiral structure of the Milky Way is not known exactly. The consensus, from the mapping of the distances to observed H$\,\textsc{II}$ regions, is that the structure can be represented by a four-arm model (e.g., \citealt{b16}; \citealt*{b15}) but the distribution of these four arms is not agreed upon. The four main arms -- Norma, Sagittarius, Perseus and Scutum--Centaurus -- are added to by the Near and Far 3-kpc arms. The Milky Way also has a central bar which can be split into a 3.1 -- 3.5 kpc Galactic Bar at an angle of 20$\degr$ with respect to the Galactic Centre-Sun axis (\citealt{b23};\citealt{b24};\citealt{b25}) and a non-axisymmetric structure, the 'Long Bar' \citep{b26}, at an angle of 44$\degr$ $\pm$ 10$\degr$ with a Galactic radius of 4.4 $\pm$ 0.5 kpc, as revealed by star counts from the $\emph{Spitzer}$ GLIMPSE survey \citep{b17}. A sketch of the structure of the Galaxy is shown in the right panel of Fig. 9 of \citet{b27}. It shows an image produced by Robert Hurt of the Spitzer Science Centre in consultation with Robert Benjamin (University of Wisconsin-Whitewater).

The region covered by this study is the slice of the Galactic Plane $\emph{l}$ = (28.5$\degr$ -- 31.5$\degr$), $\mid$\emph{b}$\mid$ $\leq$ 1$\degr$, which will hereafter be referred to as the $\emph{l}$ = 30$\degr$ region. The $\emph{l}$ = 30$\degr$ region contains the tangent of the Scutum--Centaurus arm and the near-end of the Long Bar, as well as mid-arm regions of the Sagittarius and Perseus arms in the background. The distribution of both molecular gas \citep{b4} and Herschel infrared Galactic Plane Survey (Hi-GAL) sources \citep{b34} from determined distances provide evidence for the separation of these structures. The presence of these large-scale structure features makes $\emph{l}$ = 30$\degr$ a perfect region to test the effect of Galactic structure on the star-formation process, in particular the stellar initial mass function (IMF) and SFE.

The $\emph{l}$ = 30$\degr$ region is home to the W43 giant molecular complex. Galactic Plane surveys in all wavelengths show W43 to be one of the largest and brightest star-forming structures found in the Galaxy. It houses a giant H$\,\textsc{II}$ region, which is particular luminous, emitting 10$^{51}$ Lyman continuum photons s$^{-1}$ with a far-IR luminosity of 3.6 $\times$ 10$^{6}$ L$_{\odot}$ \citep{b47}. The ionising source of this H$\,\textsc{II}$ region is a cluster of Wolf--Rayet (W--R) and OB main-sequence stars \citep*{b48} at $\emph{l}$ $\sim$ 30.75$\degr$, $\emph{b}$ $\sim$ -0.06$\degr$.

The W43 giant molecular complex is considered to be a mini-starburst region. The infrared luminosity and number of Lyman continuum photons are numbers representative of those found in starburst galaxies such as M82 \citep{b49}. W43 is also undergoing an intense episode of high-mass star formation, $\sim$ 15 high-mass protoclusters with a star formation rate (SFR) of $\sim$ 25 per cent / 10$^{6}$ yr \citep*{b50}.

We aim to measure the dense clump analogues (or precursors) of the IMF and SFE, the clump mass function and clump formation efficiency, by using the Bolocam Galactic Plane Survey (BGPS) source catalogue \citep{b10}. An estimate of the clump formation efficiency, the ratio of clump-to-cloud mass, was made by \citet{b21} of the W43 giant molecular complex. We will discuss this estimate.

We assign kinematic distances to the BGPS sources within the $\emph{l}$ = 30$\degr$ region by correlating their spatial positions and velocities with a catalogue of molecular clouds from the Galactic Ring Survey (GRS). We also use other methods for BGPS sources which cannot be associated with GRS clouds, which will be described in Sect.3. In the next section (Sect.2), we give a brief overview of the data sets used. Sect.4 contains the results and analysis and Sect.5 is the discussion of the results. Sect.6 is a summary of the conclusions.

\section{Data Sets and Observations}

\subsection[]{The BU-FCRAO Galactic Ring Survey}

The Galactic Ring Survey (GRS; \citealt{b1}) mapped $^{13}$CO $J$ = 1 $\rightarrow$ 0 emission in Galactic longitude from $\emph{l}$ = 18$\degr$ to 55.7$\degr$ and $\mid$\emph{b}$\mid$ $\leq$ 1$\degr$, covering a total area of 75.4 deg$^{2}$. The velocity range of the survey was -5 -- 135 km s$^{-1}$ for Galactic longitudes $\emph{l}$ $\leq$ 40$\degr$ and -5 -- 85 km s$^{-1}$ for $\emph{l}$ $>$ 40$\degr$ with an RMS sensitivity of $\sim$ 0.13-K. The GRS is fully sampled with a 46 arcsec angular resolution on a 22 arcsec grid and has a spectral resolution of 0.21 km s$^{-1}$.

\citet{b2} published a catalogue of 829 molecular clouds within the GRS region, identified using the CLUMPFIND algorithm \citep*{b3}. The distances to 750 of these clouds were determined by \citet{b4} using H$\,\textsc{I}$ self-absorption (HISA) to resolve the kinematic distance ambiguities. This cloud distance catalogue was complemented by the work of \citet{b6} who made use of $^{12}$CO $J$ = 1 $\rightarrow$ 0 emission from the UMSB (University of Massachusetts-Stony Brook) survey (\citealt{b7}; \citealt{b8}) to derive the masses, as well as other physical properties, of 580 molecular clouds. A power-law relation between their radii and masses was produced to allow the masses for a further 170 molecular clouds to be estimated. The associated cloud mass uncertainties are also catalogued.

The molecular cloud mass completeness limit of the GRS as a function of distance is $\emph{M$_\rmn{min}$}$ = 50$\emph{d$^{2}$}$ M$_{\odot}$, where $\emph{d}$ is the distance in kpc, with the survey complete above a mass of 4 $\times$ 10$^{4}$ M$_{\odot}$  out to the 15 kpc distance probed by the GRS \citep{b6}.

\subsection[]{The Bolocam Galactic Plane Survey}

The Bolocam Galactic Plane Survey (BGPS; \citealt{b9}) mapped 133 deg$^{2}$ of the northern Galactic Plane in the continuum at 271.1 GHz (1.1 mm). This was done with a  bandwidth of 46 GHz, an RMS noise level of 11--53 mJy beam$^{-1}$ and an effective angular resolution of 33 arcsec. The survey was continuous from $\emph{l}$ = -10.5$\degr$ to 90.5$\degr$, $\mid$\emph{b}$\mid$ $\leq$ 0.5$\degr$. There are also cross-cuts which flare out to $\mid$\emph{b}$\mid$ $\leq$ 1.5$\degr$ at $\emph{l}$ = 3$\degr$, 15$\degr$, 30$\degr$ and 31$\degr$ and towards the Cygnus X massive star forming region at $\emph{l}$ = 75.5$\degr$ to 87.5$\degr$. A further 37 deg$^{2}$ were observed towards targeted regions in the outer Galaxy, bringing the total survey area to 170 deg$^{2}$.

A custom source extraction algorithm, Bolocat, was designed and utilised to extract 8358 sources, with a catalogue 98\% complete from 0.4 -- 60 Jy over all sources with object size $\leq$ 3.5$\arcmin$. The completeness limit of the survey varies as a function of longitude, with the flux density completeness limit taken as 5 times the median RMS noise level in 1$\degr$ bins \citep{b10}. They concluded that the extracted sources were best described as molecular clumps - large, dense, bound regions within which stellar clusters and large systems form.

\subsection[]{$^{13}$CO $J$ = 3 $\rightarrow$ 2 Data}

The $\emph{l}$ = 30$\degr$ region was mapped in $^{13}$CO $J$ = 3 $\rightarrow$ 2 (330.450 GHz) with the Heterodyne Array Receiver Programme (HARP) at the James Clerk Maxwell Telescope (JCMT) on Mauna Kea, Hawaii. HARP has 16 receptors, each with a beamsize of $\sim$ 14 arcsec, separated by 30 arcsec and operates in the 325 -- 375 GHz band \citep{b32}. Observations were made in two parts, $\emph{l}$ = 30$\degr$ -- 32$\degr$ in 2008 and  $\emph{l}$ = 28$\degr$ -- 30$\degr$ in 2010. The Galactic latitude range of these observations is $\mid$\emph{b}$\mid$ $\leq$ 0.5$\degr$. The data were used in their partly reduced state for this study, as it was only necessary to identify the velocity of the peaks in the spectra extracted. The observations and reduction procedure will be discussed in more detail in a later paper.

The higher energy transition of $J$ = 3 $\rightarrow$ 2 traces higher density gas than the $J$ = 1 $\rightarrow$ 0 transition. It has a critical density of $>$$\sim$ 10$^{4}$cm$^{-3}$, compared to 10$^{2}$ -- 10$^{3}$ cm$^{-3}$ for $J$ = 1 $\rightarrow$ 0, and $\emph{E(J = 3)/k}$ = 32.8 K so is also biased towards warmer gas. $J$ = 3 $\rightarrow$ 2 is therefore less ambiguous than $J$ = 1 $\rightarrow$ 0 in identifying the emission from dense, star-forming clumps, where there are multiple emission components within a spectrum along a particular line of sight.

\subsection[]{The VLA Galactic Plane Survey}

The VLA Galactic Plane Survey (VGPS) \citep{b5} mapped H$\,\textsc{I}$ and 21-cm continuum emission in Galactic longitude from $\emph{l}$ = 18$\degr$ to 67$\degr$ and in Galactic latitude from $\mid$\emph{b}$\mid$ $\leq$ 1.3$\degr$ to $\mid$\emph{b}$\mid$ $\leq$ 2.3$\degr$. The survey was conducted with an angular resolution of 1$\arcmin$, with a spectral resolution of 1.56 km s$^{-1}$ and an RMS sensitivity of 2 K.

\section{BGPS Source Distance Determination}

\subsection[]{Matching BGPS sources with GRS clouds}

\begin{figure}
\includegraphics[width=0.5\textwidth]{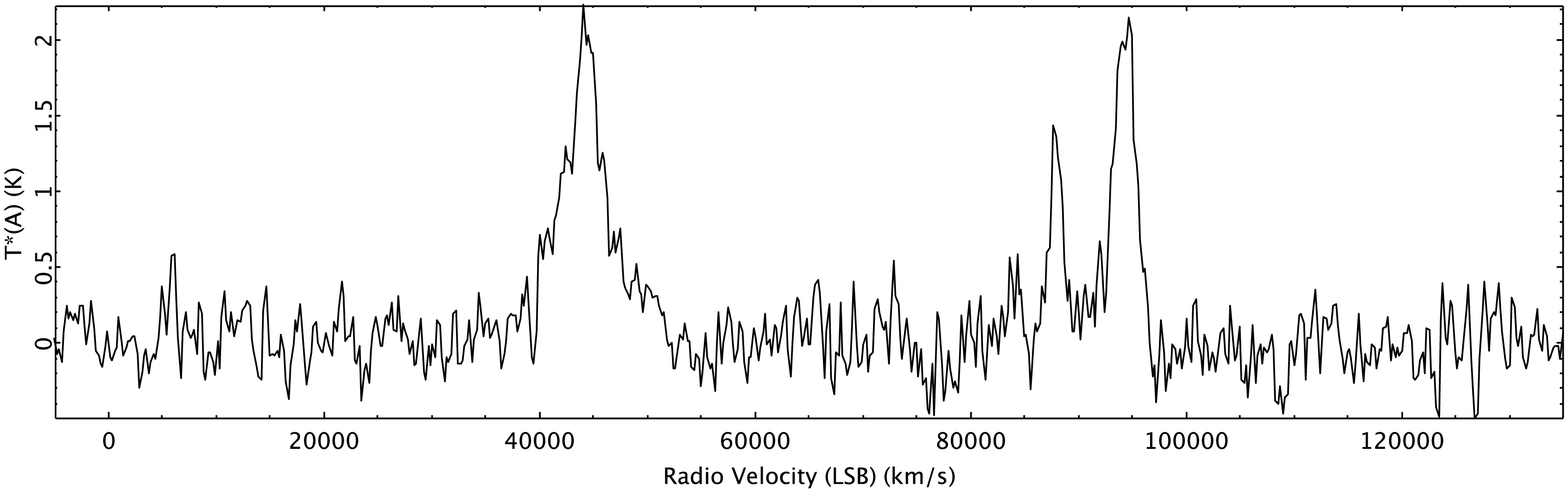}
\includegraphics[width=0.5\textwidth]{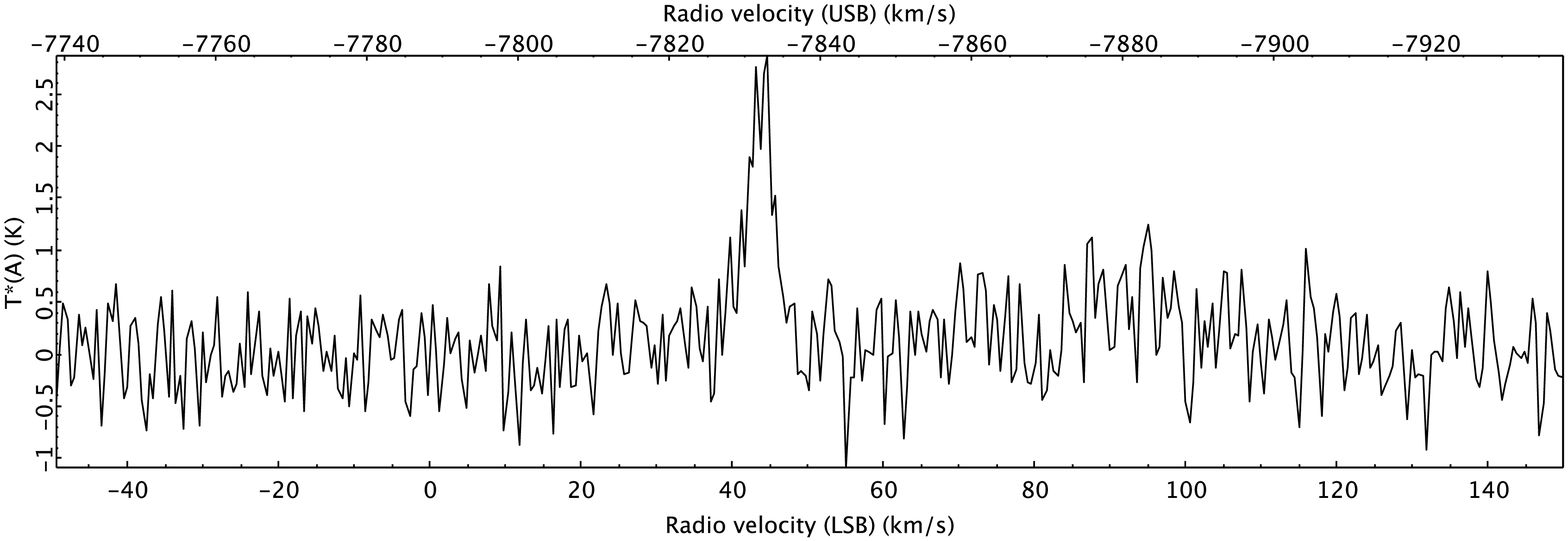}
\caption{Top panel: A spectrum from the GRS $^{13}$CO $\emph{J}$ = 1 $\rightarrow$ 0 data, displaying multiple emission components. Bottom panel: The corresponding spectrum from the HARP $\emph{J}$ = 3 $\rightarrow$ 2 data. These are the spectra of BGPS source 4423, found at $\emph{l}$ = 30.486$\degr$, $\emph{b}$ = -0.022$\degr$. This source was assigned to GRS cloud G030.59-00.01, at a distance of 11.75 kpc.}
\label{spectra}
\end{figure}

The $\emph{l}$ = 30$\degr$ region contains 806 BGPS sources and 112 GRS clouds, with distances known to 105 of these clouds \citep{b4}. We assigned kinematic velocities to the BGPS sources by extracting spectra from the GRS data cubes at the BGPS catalogue position. For sources whose spectra displayed more than one significant emission peak in the $J$ = 1 $\rightarrow$ 0 data, the higher density tracing $^{13}$CO $J$ = 3 $\rightarrow$ 2 HARP data were used. For those BGPS sources with multiple peaks in the $J$ = 3 $\rightarrow$ 2 spectra, the strongest emission feature was chosen. An example of the use of the HARP data is shown in Fig.~\ref{spectra}, with the HARP spectrum showing just one significant emission peak, whilst the GRS spectrum on the same line of sight displays multiple components.

The positions and assigned velocities of the BGPS sources were matched to those derived for the clouds in the \citet{b2} catalogue. The tolerance in position was 5 $\times$ 5 resolution elements in the $\emph{l}$ $\times$ $\emph{b}$ directions, which corresponds to 110$\arcsec$ $\times$ 110$\arcsec$, from emission associated with a cloud and the tolerance in velocity was the velocity FWHM from the aforementioned catalogue. This resulted in cloud associations for 640 BGPS sources which were then assigned distances equal to those of the corresponding GRS clouds \citet{b4}. 11 of the BGPS-GRS associations fall in the set of clouds with no known distance. These clouds had no known distance due to the disagreement between the HISA and 21-cm continuum source methods used to determine distance.

The remaining 166 BGPS detections are unassociated with GRS clouds, since no matching clouds were found but they do have GRS emission. This may be because the assigned velocities were incorrect, or because the host clouds were not identified by the GRS survey. To verify the velocities for these unmatched objects, we first concentrated on those BGPS sources whose GRS spectra displayed only one emission peak. The $J$ = 3 $\rightarrow$ 2 spectra, where available, were extracted to check if the velocity assignments were consistent with those from the $J$ = 1 $\rightarrow$ 0 data. This was found to be true in most cases, so it was unlikely that the velocity assignments were incorrect. For instances where $J$ = 3 $\rightarrow$ 2 spectra were used as the primary velocity indicator, the spectra were checked for secondary emission components and any such components were consistent with a GRS cloud. This resulted in a further 24 associations, with one of these new associations falling in a cloud with no known distance.

The spatial coincidence with emission was then tested. We produced velocity-integrated maps of the $^{13}$CO $J$ = 1 $\rightarrow$ 0 emission, and found emission coincident with 140 of the 142 remaining BGPS sources. This emission arises from either relatively bright, compact regions, some with very small velocity ranges or from filamentary-type clouds. These small, low velocity-dispersion clouds likely fell below the detection criteria used by \citet{b2} of $\Delta\emph{l}$  or $\Delta\emph{b}$ $\leq$ 6$\arcmin$ or $\Delta$V $\leq$ 0.6 km s$^{-1}$, with the filamentary clouds probably falling below the $\Delta\emph{l}$  or $\Delta\emph{b}$ criterion. We found that 74 BGPS sources are coincident with 56 small, low velocity-dispersion clouds, with 66 BGPS sources falling in 33 filamentary-type clouds.

Of the 142 sources not associated with GRS clouds, one fell outside of the latitude range of the GRS data (BGPS source 4308 at $\emph{b}$ = -1.15$\degr$), so no kinematic velocity was able to be assigned.

Table~\ref{assocs} displays a summary of the GRS cloud-BGPS source associations, displaying the associations for individual clouds (only a small portion of the data is provided here, with the full list of 664 BGPS sources available as Supporting Information to the online article).

\begin{table}
\begin{center}
\caption{Summary of GRS cloud parameters and BGPS source associations. Only a small portion of the data is provided here. The full list of 664 BGPS sources is available as Supporting Information to the online article.}
\label{assocs}
\begin{tabular}{lccccc} \hline
 GRS Cloud &  GRS & GRS & BGPS & BGPS & BGPS \\
Name & $\emph{V$_{LSR}$}$ & $\emph{D}$ & Source & $\emph{l}$ & $\emph{b}$ \\
 & (km s$^{-1}$) & (kpc) & ID & ($\degr$) & ($\degr$) \\
 \hline
 G028.34+00.04 & 78.57 & 5.00 & 3984 & 28.503 & 0.290 \\
 G028.34+00.04 & 78.57 & 5.00 & 3986 & 28.505 & 0.106 \\
 G028.34+00.04 & 78.57 & 5.00 & 3994 & 28.533 & 0.128 \\
 G028.34+00.04 & 78.57 & 5.00 & 3996 & 28.547 & 0.064 \\
 G028.34+00.04 & 78.57 & 5.00 & 4001 & 28.581 & 0.144 \\
 G028.34+00.04 & 78.57 & 5.00 & 4005 & 28.603 & -0.082 \\
 G028.34+00.04 & 78.57 & 5.00 & 4008 & 28.619 & -0.070 \\
 G028.34+00.04 & 78.57 & 5.00 & 4009 & 28.629 & 0.158 \\
 G028.34+00.04 & 78.57 & 5.00 & 4015 & 28.659 & 0.144 \\
 G028.39-00.01 & 98.98 & 6.32	 & 3985 & 28.505 & -0.142 \\
 G028.39-00.01 & 98.98 & 6.32 & 3989 & 28.507 & -0.176 \\
 G028.39-00.01 & 98.98 & 6.32 & 3993 & 28.525 & 0.070 \\
 G028.39-00.01 & 98.98 &6.32 & 4032 & 28.759 & 0.106 \\
 G028.49+00.19 & 101.10	 & 6.53 & 3987 & 28.505 & 0.196 \\
 G028.49+00.19 & 101.10	 & 6.53 & 3991 & 28.513 & 0.240 \\
 \hline
 \end{tabular}
 \end{center}
 \end{table}

\subsection[]{Distance determination of remaining BGPS sources}

\subsubsection[]{Determining kinematic distances from Galactic rotation curve}

\begin{figure}
\includegraphics[scale=0.5]{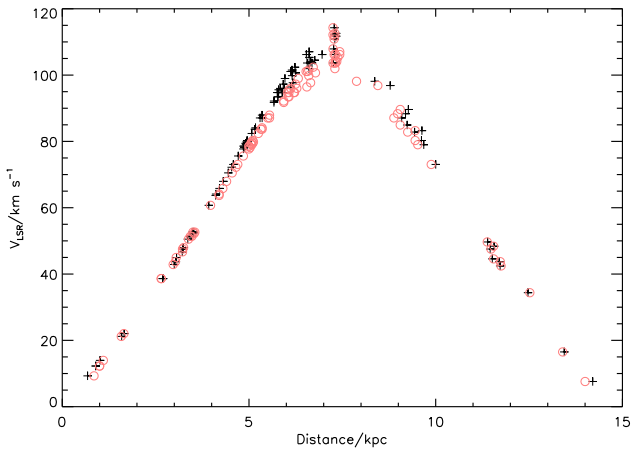}
\caption{Comparison of calculated distances to GRS clouds using the \citet{b29} rotation curve, indicated by the black, cross-shaped symbols, and the kinematic distances calculated using the \citet{b30} rotation curve, indicated by the red, circular symbols.}
\label{rotation}
\end{figure}

Each of the remaining 141 unassociated BGPS sources had two kinematic distances calculated by using its assigned velocity and the Galactic rotation curve of \citet{b30}. This Galactic rotation curve was favoured over \citet{b29} and \citet{b31} because this particular rotation curve makes use of data from all four Galactic quadrants, and is therefore more representative of the general rotation of the Galaxy. However, the discrepancies are small. Fig.~\ref{rotation} shows a comparison between the kinematic distances of the GRS clouds calculated using the \citet{b29} rotation curve and the kinematic distances calculated using the \citet{b30} rotation curve. The plot shows that the two rotation curves produce very similar results, with the most significant departures of $\sim$ 0.9 kpc occurring at distances of $\sim$ 8 kpc.

The kinematic distances were then calculated from the \citet{b30} rotation curve, with all constants coming from within \citet{b30}, by first deriving an angular velocity from the line-of-sight velocity, $\emph{V}_{LSR}$, via the relation:

\begin{equation}
\omega = \omega_{0} + \frac{V_{LSR}}{R_{0}\rmn{sin}(\emph{l})\rmn{cos}(\emph{b})}
\end{equation}

\noindent where $\omega_{0}$ and $\emph{R}_{0}$ are the Solar angular velocity and galactocentric distance with assumed values of 220 km s$^{-1}$ and 8.5 kpc respectively.

The rotation curve is then numerically inverted to give $\emph{R}$, with the rotation curve being:

\begin{equation}
\frac{\omega}{\omega_{0}} = a_{1}\left(\frac{R}{R_{0}}\right)^{a_{2}-1} + a_{3}\left(\frac{R_{0}}{R}\right)
\end{equation}

\noindent where $\emph{R}$ is the derived galactocentric distance and $\emph{a}_{1}$, $\emph{a}_{2}$ and $\emph{a}_{3}$ are constants with values of 1.0077, 0.0394 and 0.0071 respectively. This allows two kinematic distances to be calculated from the roots of:

\begin{equation}
d^{2}\rmn{cos}^{2}(\emph{b}) - (2R_{0}\rmn{cos}(\emph{b})\rmn{cos}({\emph{l}))d + (R_{0}^2} - R^{2}) = 0
\label{dist}
\end{equation}

The unassociated BGPS sources were checked to see which could be grouped together as coming from the same source of GRS emission via their spatial location and assigned velocity. This was important for those sources close to the tangent point. The tangent point is where the near and far kinematic distances are equal. Sources in or around the tangent point can have undetermined kinematic distances as the determinant of equation~\ref{dist} may fall below zero. In these instances, the distance is calculated to be:

\begin{equation}
d = \frac{R_{0}\rmn{cos}(\emph{l})}{\rmn{cos}(\emph{b})}
\end{equation}

\subsubsection[]{Determining between two kinematic distances}

The first technique used to determine which kinematic distance should be used is a scale height cutoff. Following \citet{b27}, a cutoff of 120 pc was used, which corresponds to 4 times the Galactic scale height of $\sim$ 30 pc measured by \cite{b35} from the distribution of nearby main sequence OB stars. Where the height above the plane of any source at its far kinematic distance is greater than this, it is unlikely that it is located at that distance, so the near distance is adopted. This applied to 6 BGPS sources.

The remaining BGPS sources  without associations to clouds in the GRS catalogue were assigned distances by using the H$\,\textsc{I}$ self-absorption (HISA) method (e.g. \citealt{b36}; \citealt{b4}). The near and far kinematic distances have the same radial velocity, as discussed above. H$\,\textsc{I}$ embedded in the cold central regions of molecular clouds (e.g. \citealt*{b37}) is much cooler ($\emph{T}$ = 10 -- 30 K) than the warm H$\,\textsc{I}$ which permeates the interstellar medium (ISM, $\emph{T}$ = 100 -- 10$^{4}$ K). If the molecular cloud is at the near distance, the cold foreground H$\,\textsc{I}$ will absorb the emission from the warm, background H$\,\textsc{I}$ and produce an absorption feature in the H$\,\textsc{I}$ spectrum which is coincident in velocity with a $^{13}$CO emission line. If the cold H$\,\textsc{I}$ is in the background, it will not absorb any warm H$\,\textsc{I}$ emission from the Galactic plane, and the H$\,\textsc{I}$ spectrum will show no absorption feature. 30 BGPS sources were assigned the far kinematic distance, with 105 found to be at the near distance due to the presence of HISA. Table~\ref{unassoc} displays a summary of the unassociated sources and their derived kinematic distances (only a small portion of the data is provided here, with the full list of 141 BGPS sources available as Supporting Information to the online article).

\begin{table}
\begin{center}
\caption{Summary of unassociated BGPS sources and their derived kinematic distances. Only a small portion of the data is provided here. The full list of 141 sources is available as Supporting Information to the online article.}
\label{unassoc}
\begin{tabular}{lccccc} \hline
BGPS Source & $\emph{l}$ & $\emph{b}$ & $\emph{V$_{LSR}$}$ & $\emph{D}$ & Scale Height\\
ID & ($\degr$) & ($\degr$) & (km s$^{-1}$) & (kpc) & Selected \\
 \hline
4513	 & 30.71 & -0.32 & 100.49	 & 6.38 & \\
4521	 & 30.73 & 0.38 & 81.97 & 9.53 & \\
4524	 & 30.74 & 0.38	 & 45.19 & 3.07	 & \\
4528 & 30.75 & 0.19	 & 73.09 & 4.59	 & \\
4536	 & 30.77 & -0.78 & 77.94 & 4.86 & y \\
4538 & 30.77 & -0.33 & 105.97	 & 7.30 & \\
4540	 & 30.77 & 0.39	 & 86.54 & 9.26 & \\
4545	 & 30.78 & 0.28	 & 42.30 & 2.90	 & \\
4559 & 30.82 & -0.31 & 89.20 & 5.52 & \\
4584 & 30.88 & -0.09 & 101.32 & 8.07 & \\
4592	 & 30.89 & -0.15 & 106.20 & 7.29 & \\
4602	 & 30.90 & -0.03 & 74.58 & 9.91 & \\
4610	 & 30.93 & -0.17 & 106.84	 & 7.29 & \\
4612 & 30.94 & -0.19 & 88.99 & 5.51 & \\
4623 & 30.95 & -0.39 & 85.38 & 5.29 & \\
\hline
\end{tabular}
\end{center}
\end{table}

\subsection[]{Summary of distance determinations}

The association of BGPS sources with GRS clouds resulted in  664 matches, with 652 having known distances. The use of velocities from CO emission not associated with recognised clouds and the rotation curve of \cite{b30} resulted in 141 further distance determinations. The breakdown of this 141 was as follows: 6 using the scale height cutoff and 135 using HISA with 30 assigned the far kinematic distance and the remaining 105, the near distance. One source fell outside the latitude range of the molecular data, and therefore had no kinematic velocity assigned to it and, as a result, no kinematic distance. This left a total of 13 out of 806 BGPS sources without a derived kinematic distance.

There are associated errors with these distance determinations. A full discussion of the distance determinations involved with the GRS clouds and velocity assignments can be found in \citet{b4}.

\subsection[]{BGPS-GRS cloud associations}

We have identified the molecular clouds associated with 664 BGPS sources. In Table~\ref{total} we present a summary of the GRS clouds with the number of associated BGPS (only a small portion of the data is provided here, with the full list of 112 clouds available as Supporting Information to the online article). We found that these BGPS sources are associated with 87 of the 112 GRS clouds from the catalogue of \citet{b2} in the $\emph{l}$ = 30$\degr$ region, with 10 of those 87 clouds having just one BGPS source associated with it.

\begin{table*}
\begin{center}
\caption{Summary of GRS cloud parameters, number of BGPS source associations and the associated BGPS source masses. Only a small portion of the data is provided here. The full list of 112 clouds is available as Supporting Information to the online article.}
\label{total}
\begin{tabular}{lccccccc} \hline
 GRS Cloud & $\emph{l}$ & $\emph{b}$ & $\emph{V$_{LSR}$}$ & $\emph{M$_{cloud}$}$ & $\emph{D}$ & No. BGPS & $\emph{$M_{clumps}$}$\\
Name & ($\degr$) & ($\degr$) & (km s$^{-1}$) & (M$_{\odot}$) & (kpc) & Sources & (M$_{\odot}$)\\
 \hline
G028.34-00.46	& 28.34 &	-0.46	 & 42.87 & 6520 & 2.97 &	0 & 0\\
G028.34-00.41	& 28.34 &	-0.41	& 84.90 & - - - & - - - & 0 &	 - - -\\
G028.34-00.06	& 28.34 &	-0.06	 & 34.37 &	 35300 & 12.52	& 0 & 0\\
G028.34-00.01	& 28.34 &	-0.01 & 46.69 &	 6940 & 3.22 &	0 & 0\\
G028.34+00.04 & 28.34 & 0.04	 & 78.57 & 634000 & 5.00 & 9 & 1770\\
G028.39-00.41	& 28.39 &	-0.41 & 73.05 & 106000 &	 4.70 & 0 & 0\\
G028.39-00.01	& 28.39 &	-0.01 & 98.98 &	 142000 &	 6.32	& 4 & 1710\\
G028.44-00.01	& 28.44 &	-0.01 & 43.72 &	 1030 & 3.03 &	0 & 0	\\
G028.49+00.19 & 28.49 & 0.19	 & 101.10 & 39000 & 6.53 & 2 & 378\\
G028.49+00.34 & 28.49 & 0.34 & 38.60 & - - - & - - - & 0 & - - -\\
G028.59-00.26	& 28.59 &	-0.26 & 63.70 & 4100 & 4.20 & 0 & 0\\
G028.59-00.21	& 28.59 &	-0.21 & 87.08 & 64500 & 5.50 & 9 &	9770\\
G028.59+00.04 & 28.59 & 0.04 & 101.53 & 82100 & 6.60	& 5 & 8940\\
G028.64+00.09 & 28.64 & 0.09	 & 107.05 & 50000 &	7.43	& 5 & 8510\\
G028.69+00.04 & 28.69 & 0.04 & 106.20 & 62200 & 7.43 & 1 & 960\\
G028.74-00.21	& 28.74 &	-0.21 & 102.38	& 28000 & 6.72 & 2 & 1040\\
G028.74+00.44 & 28.74 & 0.44 & 92.20 & - - - & - - - & 4 &  - - - \\
G028.84-00.36	& 28.84 &	-0.36 & 101.10 & 28700 & 6.57 & 2 & 509\\
G028.84-00.26	& 28.84 &	-0.26 & 87.93 & 44800 & 5.55 & 7 &	7380\\
G028.84+00.14 & 28.84 & 0.14	 & 52.22 & 2230 & 3.53 & 7 & 405\\
\hline
\end{tabular}
\end{center}
\end{table*}
 
\begin{figure}
\begin{center}
\includegraphics[scale=0.5]{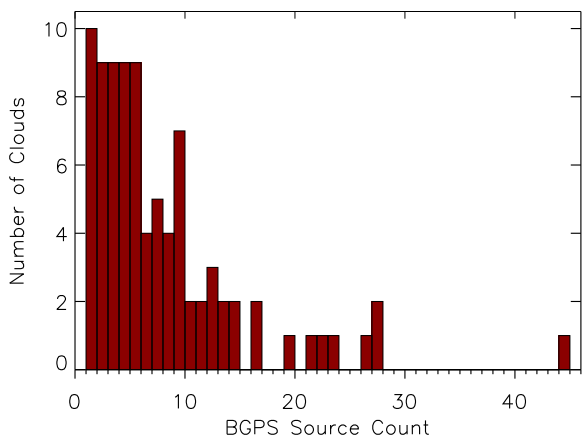}
\caption{The number of BGPS sources found in each molecular cloud with at least one association.}
\label{distri}
\end{center}
\end{figure}

In Fig.~\ref{distri} we present a histogram of the number of BGPS sources found in each GRS molecular cloud. The majority of the clouds are each associated with less than 10 clumps. However, there are 7 clouds which are associated with more than 20 BGPS sources each. 6 of these clouds could be considered part of the W43 giant molecular complex \citep{b21}, whilst one cloud (G031.24-00.01; \citealt{b2}) is found to be in the Perseus spiral arm which runs across the $\emph{l}$ = 30$\degr$ field.

\section{Results \& Analysis}

\subsection[]{Populations within the $\emph{l}$ = 30$\degr$ region}

The $\emph{l}$ = 30$\degr$ region is host to major Grand Design features of the Galaxy. It contains the tangent of the Scutum--Centaurus arm and the near end of the Long Bar (which together will hereafter be referred to as the tangent region), as well as mid-arm sections of the Sagittarius and Perseus arms in the background. There is also structure in the foreground, the nature of which is undetermined but is possibly the near-part of the Sagittarius arm.

In aiming to test the effect of Galactic structure on the star-formation process, it is first important to try to distinguish between these three regions: the foreground spur/Sagittarius arm; the background Perseus and Sagittarius arms and the tangent region, which dominates the emission from this field as we are looking directly down the Scutum--Centaurus arm. The separation of the three components is not an exact process, but the main aim is to ensure that there is sufficient separation such that emission in each of the three distance bins is dominated by sources within the main structure in each case.

The W43 star-forming complex is thought to lie on the tangent point \citep{b21} and as such, dominates the emission of the entire region. This allows us to conclude that if emission can be associated with W43, it can also be associated with the tangent region. Galactic rotation curves give two kinematic distances for W43, a near distance of $\sim$ 5.9 kpc and a far distance of $\sim$ 8.7 kpc \citep{b21}. Most authors have argued the near distance with \citet{b36} using HISA, \citet*{b38} using near-infrared extinction and \citet{b39} using H$\,\textsc{II}$ regions. The Herschel infrared Galactic Plane Survey (Hi-GAL) distance determination also assumes the near kinematic distance \citep{b34}. We assume the near kinematic distance due to the distance distribution of the GRS clouds from the \citet{b4} catalogue.

\begin{figure}
\begin{center}
\includegraphics[scale=0.5]{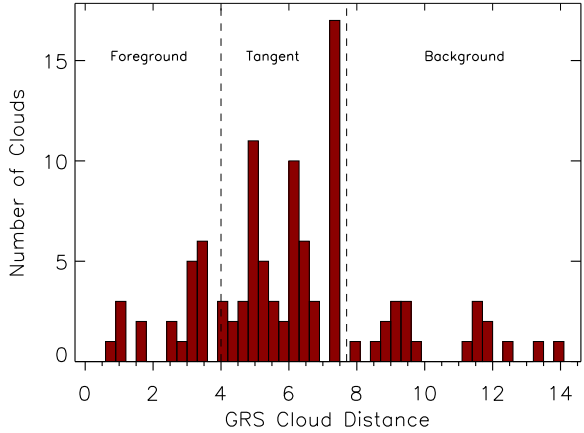}
\caption{The number of GRS clouds found as a function of distance. The vertical dashed lines indicate the chosen boundaries of the tangent region.}
\label{distance}
\end{center}
\end{figure}

Fig~\ref{distance} shows that the clouds associated with the tangent (and W43) are clustered in distance between $\sim$ 4.0 kpc and $\sim$ 7.7 kpc and so this was taken to be the distance bin defining the tangent region. All the GRS clouds and BGPS sources within this range were considered to be associated with the tangent region, while clouds with distances less than 3.9 kpc were considered foreground, and greater than 7.7 kpc were considered background. There is evidence in Fig~\ref{distance} for the two background arms with the Sagittarius arm at $\sim$ 9 kpc and the Perseus arm at $\sim$ 11 kpc. This distance to the Perseus arm is supported by the work of Summers et al. (2012, in prep), whose models of the distribution of the molecular gas component of the Galaxy found the arm at a distance of 11.8 kpc along this line of sight.

\subsection[]{Masses of BGPS sources}

\begin{figure}
\begin{center}
\includegraphics[scale=0.5]{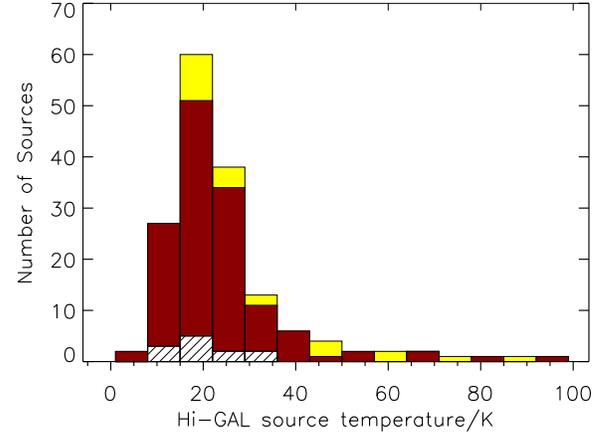}
\caption{The distributions of SED-based temperatures of the Hi-GAL SDP sources for the $\emph{l}$ = 30$\degr$ region with the tangent, background and foreground components depicted by the red, yellow and white hashed bars respectively. The regions were separated using the \citet{b34} distances.}
\label{Hi-GAL}
\end{center}
\end{figure}

The 1.1-mm dust continuum flux densities from BGPS are converted to source masses using the standard formula:

\begin{equation}
M = \frac{S_{\nu}D^{2}}{\kappa_{\nu}B_{\nu}(T_{d})}
\end{equation}

\noindent which leads to:

\begin{equation}
M=13.1 M_{\odot} \left(\frac{S_{\nu}}{1 Jy}\right)\left(\frac{D}{1 kpc}\right)^{2}(\rmn{e}^{13.12/T_{d}}-1)
\end{equation}

\noindent where we assume $\kappa_{\nu}$ = 0.0114 cm$^{2}$ g$^{-1}$ \citep{b40}, $\emph{S$_{\nu}$}$ is the catalogued flux density, $\emph{D}$ is the source distance and $\emph{B$_{\nu}$}$ is the Planck function evaluated at dust temperature, $\emph{T$_{d}$}$. The value for the opacity is obtained by using a dust temperature of 20 K and the dust opacities of \citet{b41}, as described by \citet{b40}. As a result, we use a flat temperature of 20 K for all the source temperatures. This choice is supported by the distribution of temperatures of the Herschel infrared Galactic Plane Survey (Hi-GAL), a $\emph{Herschel Space Observatory}$ Open-time Key Project, Science Demonstration Phase (SDP) \citep{b28} sources for the $\emph{l}$ = 30$\degr$ region. The $\emph{l}$ = 30$\degr$ region was chosen as one of the two SDP fields. The temperatures were derived by SED and greybody fitting \citep{b42} and Fig.~\ref{Hi-GAL} shows that the peak of the distribution lies at 20 K in all 3 components of the  $\emph{l}$ = 30$\degr$ region.

The flux densities for each source also require a further multiplication by a factor of 1.5 to provide consistency with other data sets from MAMBO and SIMBA surveys \citep{b9}.

The choice of temperature is not the important factor in this study, as the premise is to compare the star-formation properties across the three regions. A more important consideration is whether the temperatures can be assumed to be the same in the three regions. Using the SED-based temperatures of the Hi-GAL SDP sources in the $\emph{l}$ = 30$\degr$ region, a Kolomogorov--Smirnov (K--S) test was applied to the $\emph{T}_{d}$ distributions of the three sub-samples. The three K--S tests showed that they could be assumed to have the same temperatures as there was a 70 per cent probability that the temperatures were from the same sample. Only 85 out of the 312 sources plotted in Fig.~\ref{Hi-GAL} can be associated with BGPS sources.

It is also worth noting that absolute masses and accurate distances are not vital to most of the results in this study, especially the clump formation efficiencies (Sect 4.4). This ratio of clump-to-cloud mass has the distance dependencies, and uncertainties, cancelled out if you assume that the gas and dust are at the same distance. This analysis mainly deals with relative masses and the separation into the broad sets associated with the large-scale structure, which could be achieved with source velocities alone.

\subsection[]{Clump Mass Functions}

The clump mass function (CMF) is the frequency distribution of the masses of the dense clumps within the molecular clouds. Studies of nearby low-mass star-forming regions (e.g. \citealt*{b43}; \citealt{b44}; \citealt*{b45}) and of high-mass star-forming regions (e.g. \citealt*{b52}; \citealt{b53}; \citealt{b54}) have shown good agreement with the shape of the high-mass end of the stellar IMF. If the CMF does indeed mimic the stellar IMF, then any change in the CMF with regards to large-scale Galactic structure could signify a change in the star formation process.

\begin{figure}
\begin{center}
\begin{tabular}{c}
\includegraphics[scale=0.5]{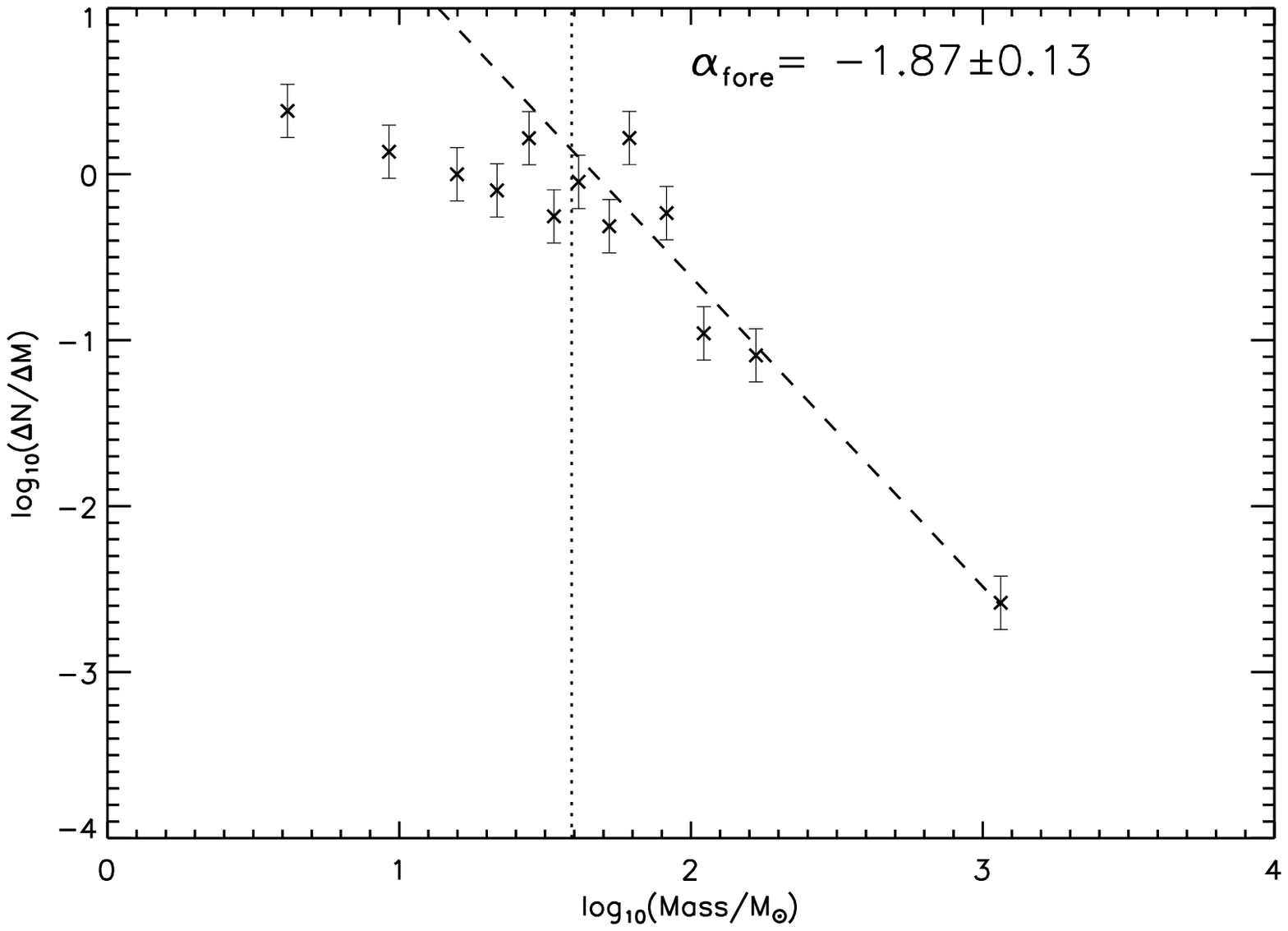}\\
\includegraphics[scale=0.5]{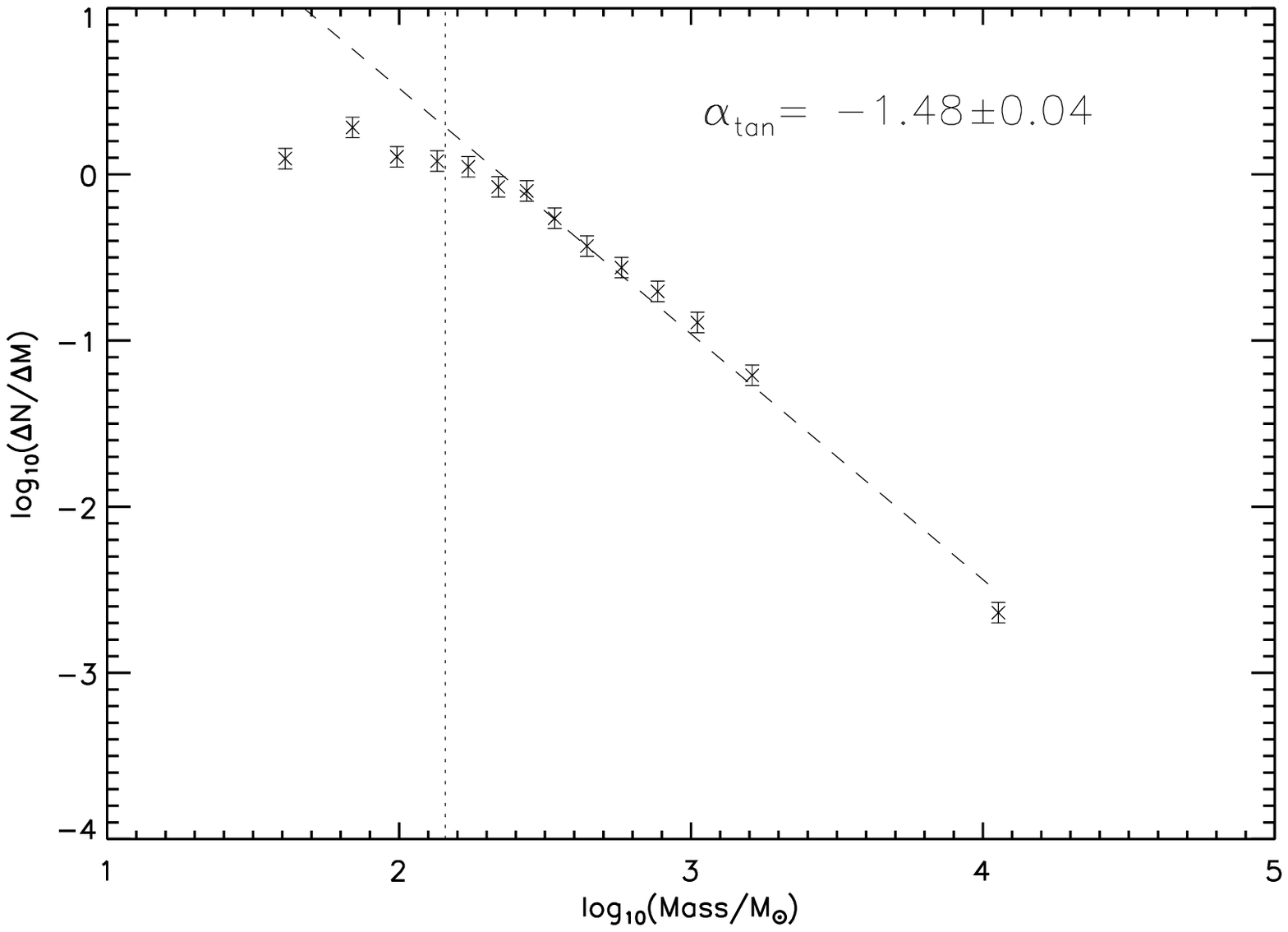}\\
\includegraphics[scale=0.5]{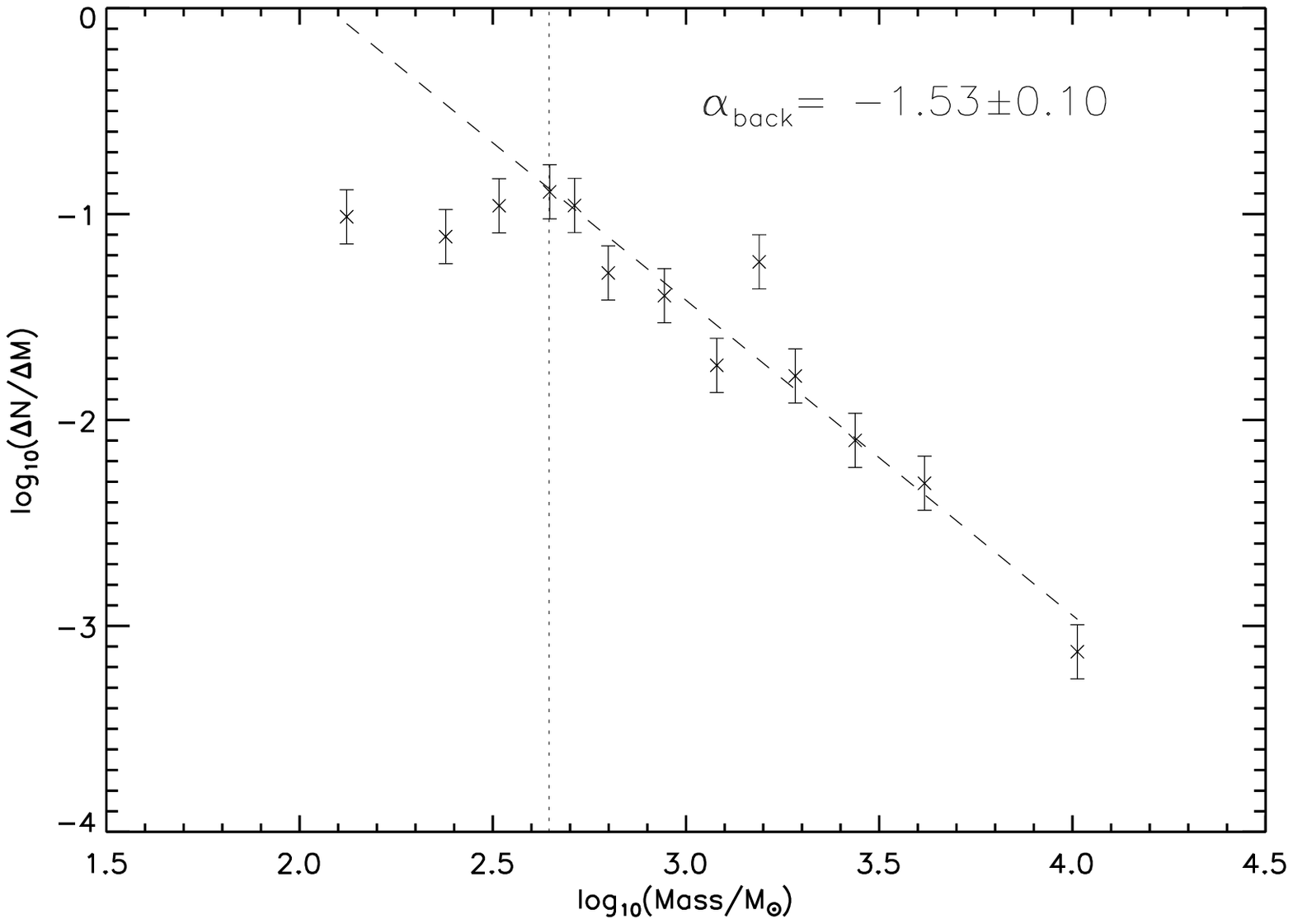}\\
\end{tabular}
\caption{CMFs for the 3 regions within the $\emph{l}$ = 30$\degr$ field. The vertical dotted lines represent the completeness limit at the furthest distance of each region. The dashed lines are fits to the slope of the CMF above the completeness limit. The error bars correspond to $\sqrt{N}$ counting statistics. Top panel: the foreground region. Middle panel: the tangent region. Bottom panel: the background region.}
\label{CMF}
\end{center}
\end{figure}

We use the masses derived above to plot CMFs of the three regions, which are presented in Fig.~\ref{CMF}. Each CMF has an associated completeness limit which corresponds to the mass detection limit at the furthest point of each velocity bin. This is calculated from the 0.135-Jy completeness limit in the BGPS 1.1-mm flux density for $\emph{l}$ = 30$\degr$ \citep{b10}. The corresponding mass limits are $\sim$ 40, 145 and 450 M$_{\odot}$ for the foreground, tangent and background regions respectively. The plotted quantity, $\Delta$$\emph{N/}$$\Delta$$\emph{M}$, is the number of clumps per unit mass interval in each mass bin. We have used a fixed number of clumps per bin, as opposed to fixed bin widths to equalise weights.

Assuming the CMF above the mass completeness limit to be a power law of the form $\Delta$$\emph{N/}$$\Delta$$\emph{M}$ $\propto$ $\emph{M}^{\alpha}$, a (least squares) fit to each CMF gives indices of  $\alpha$ = -1.87 $\pm$ 0.13, -1.48 $\pm$ 0.04 and -1.53 $\pm$ 0.10 for the foreground, tangent and background regions respectively. The index values for the background and tangent regions are consistent with each other, however the foreground region is found to depart at the 3--$\sigma$ level. This is potential evidence for a steeper slope in the foreground clumps. These slopes are more consistent with the slopes derived from gas emission \citep{b55}, as opposed to those found in the fellow dust emission studies listed above.

The blending of clumps with distance is not observed to be significant within these CMFs. If this were to be having an effect, a second turnover would be observed at masses greater than the completeness turnover. This is caused by a combination of clustering scale and angular resolution, so at some combination of the two, low-mass clumps are bumped up into the higher bin. This effect is seen in the observations of \citet{b20} and the simulations of \citet{b66}. In any case, the high-mass end slope is preserved because the most massive clumps are rare and only blend with smaller clumps \citep{b20}.

\subsection[]{Clump Formation Efficiencies}

The clump formation efficiency (CFE) is a measure of what fraction of molecular gas has been converted into dense clumps. This quantity is analogous (or a precursor) to the star formation efficiency. The CFE must be viewed as an upper limit to the SFE.

The CFE is a measure of: 

\begin{equation}
\frac{M_{clump}}{M_{cloud}}= \frac{1}{M_{cloud}} \int^t_0 \frac{dM}{dt}\,dt
\end{equation}

\noindent where $\emph{dM/dt}$ is the instantaneous clump formation rate. A high value for the CFE can indicate either a high clump formation rate or a long formation timescale.

By using the catalogued GRS cloud masses \citep{b6} and the derived masses for the BGPS sources, we are able to calculate total CFEs for the three regions of the $\emph{l}$ = 30$\degr$ field. The values obtained are 13.00 $\pm$ 1.60, 8.38 $\pm$ 0.30 and 8.52 $\pm$ 0.73 per cent for the foreground, tangent and background regions respectively. These values do not include the masses of the small, low-velocity clouds discussed in Sect 3.1 that were not included by \citet{b2}, and as such the CFEs can be taken as an upper limit. The uncertainties on the CFEs come from the catalogued GRS cloud mass uncertainties \citep{b6}, the uncertainties in BGPS flux densities \citep{b10} and the distribution of source temperatures \citep{b42}. Extrapolation of the BGPS traced mass within the background region to the completeness limit of the tangent region, to adjust for the affect of distance on detected clump masses, shows little variation with the extrapolated CFE falling within current uncertainties. The biases corresponding to the distance distribution of cloud masses are discussed in detail by \citet{b6}. The values for the CFEs in the background and tangent regions do not depart significantly from each other. However, the foreground value is higher at the 3--$\sigma$ level. It might be expected that the tangent region would have the higher CFE, due to its position at the end of the Galactic Bar and the intense star-formation associated with W43, but we find no discernible difference between it and the background spiral-arm segments crossing the field.

\citet{b21} calculate a CFE of 20 per cent for the entire W43 star-forming complex, including all the material in the $\emph{l}$ = 30$\degr$ Hi-GAL SDP field between LSR velocities of 80 and 110 km s$^{-1}$, using clump masses from ATLASGAL (The APEX Telescope Large Area Survey of the GALaxy; \citealt{b46}) and integrated $^{13}$CO $J$ = 1 $\rightarrow$ 0 emission from GRS data. While this is significantly higher than our value of $\sim$ 8.4 per cent, the difference is mostly accounted for since they did not include an optical depth correction in their molecular mass estimate. The median $^{13}$CO $J$ = 1 $\rightarrow$ 0 optical depth of clouds in GRS is around 1.3 \citep{b6}, which implies a correction factor of $\times$ 1.8 for the molecular mass, reducing the CFE of \citet{b21} to around 11 per cent. ATLASGAL has a detection limit of $\sim$ 50 M$_{\odot}$ lower than the BGPS (i.e. $\sim$ 95 M$_{\odot}$) at a distance corresponding to the furthest point of the tangent region \citep{b46} and so detects more mass in dense clumps. However extrapolating the CMF back to the ATLASGAL completeness only added 1 per cent to the CFE and there is an additional discrepancy of $\sim$ 50 per cent in the dense integrated core mass (8.4 $\times$ 10$^{5}$ M$_{\odot}$ compared to 5.4 $\times$ M$_{\odot}$) which is unidentified because we used compatible emissivity and the same dust temperature.

The association of molecular clouds with specific regions in \citet{b21} was done somewhat differently to the current study. Whereas we have used distance selection of GRS clouds from the catalogue of \citet{b4}, they used velocity selection of data direct from the GRS database. The two samples are 90 percent consistent, however.

\section{Discussion}

\subsection[]{Clump Formation Efficiency within the tangent region}

\begin{figure*}
\begin{center}
\includegraphics[scale=0.6]{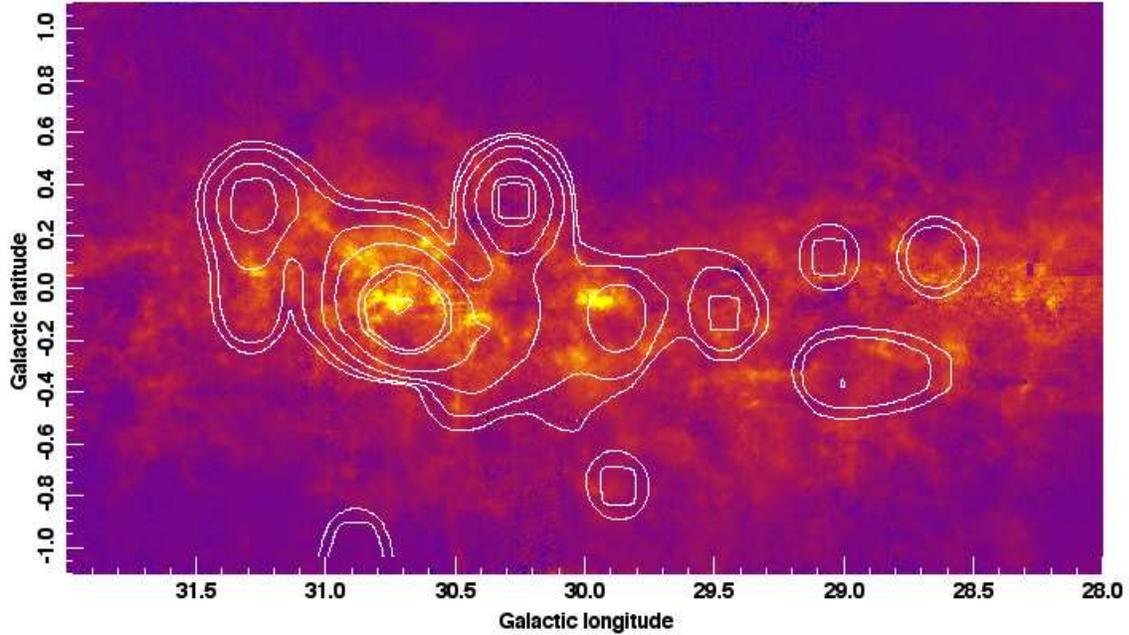}
\caption{The GRS $\emph{J}$ = 1 $\rightarrow$ 0 integrated emission of the $\emph{l}$ = 30$\degr$ region with overlaid contours of clump formation efficiencies. The contour levels are at 1.50, 3.60, 5.25, 7.50, 15.00, 22.50 and 55.00 per cent.}
\label{CFE}
\end{center}
\end{figure*}

Fig.~\ref{CFE} shows contours of CFE in individual GRS clouds overlaid on an image of integrated $^{13}$CO $J$ = 1 $\rightarrow$ 0 emission in the $\emph{l}$ = 30$\degr$ region. The CFE data have a spatial resolution of 0.2$\degr$ $\times$ 0.2$\degr$, which is consistent with the cloud size scale at the distance of the tangent clouds. The median diameter of clouds in the tangent subset is $\sim$ 20.5 pc, which corresponds to 0.3$\degr$ at 4 kpc and 0.15$\degr$ at 7.7 kpc. The CFE values are the ratio of the integrated BGPS dense-clump mass to the integrated GRS cloud mass in each spatial element. The ratios do not include the unassociated BGPS sources or the unquantified number of small, low-velocity dispersion clouds which fall below the detection thresholds of \citet{b2}. The CFE is clearly highly variable from cloud to cloud, ranging from zero to $\sim$ 60 per cent. 92 per cent of clouds with associated BGPS sources have CFE ratios of $>$0 -- 15 per cent (Fig.~\ref{CFEhist}). The highest value ($\sim$ 60 per cent) coincides with the position of the ionising source of the W43 H$\,\textsc{II}$ region which has been associated with the triggering of star formation in its vicinity \citep{b47} and the star formation in this cloud is clearly very efficient in the current epoch. The presence of a triggering agent has been observed to increase the CFE locally in the W3 giant molecular cloud \citep{b20} and such increases are predicted by simulations \citep*{b63}. The other individual star-forming regions in the field have lower CFE and this may be because the local triggering is weaker in these sources. For example, there is a localised maximum of $\sim$ 15 per cent  coincident with the H$\,\textsc{II}$ G29.96-0.02 \citep{b38}. There is also a peak of $\sim$ 7.50 per cent at $\emph{l}$ $\approx$ 31.3,  $\emph{b}$ $\approx$ 0.3 but there are very few catalogued sources in this region other than the BGPS sources. 

When quantifying the CFE (or SFE), the result will depend on the chosen size scale. It is therefore important to select a scale that has some physical meaning. The scale corresponding to the projected molecular cloud sizes is the first of these. The median projected separation of clouds is 0.15$\degr$, about the same as the cloud size scale so the next significant scale is the width of the entire molecular complex associated with the Scutum-Centarus/Long Bar tangent, in which the mean properties of the clouds may be affected by the environment in that unique location.  Below we compare the CFE values found in the tangent clouds to those in the foreground and background clouds.

\subsection[]{W43: a mini-starburst region?}

Since starbursts are associated with high star-formation efficiencies, if the W43 complex were to be considered a mini-starburst region, it might be expected to have an unusually high CFE across a wide area, as a result of special environmental conditions. We can examine this possibility. While the CFE of the individual molecular cloud containing the W43 H$\,\textsc{II}$ region is indeed very high, the median CFE values for clouds within the foreground, background and tangent regions, as defined in Sect.4.1, are found to be 7.15 $\pm$ 4.95, 3.78 $\pm$ 3.32 and 3.27 $\pm$ 2.28 per cent respectively (the uncertainties quoted here are the median absolute deviations), with the mean CFE values found to be 12.31 $\pm$ 4.86, 5.95 $\pm$ 2.26 and 5.38 $\pm$ 1.22 per cent. Since these are consistent within the uncertainties at an approximate 1--$\sigma$ level, it implies that the clouds within the tangent region are not, on average, in a significantly different physical state to those in the other two velocity bins.

The lack of difference in the CFEs and CMFs between the tangent and background regions is supported by the work of \citet{b56}. They find that the regions of dense gas traced by the BGPS show little variation across the Galaxy, regardless of environment.

\begin{figure}
\begin{center}
\includegraphics[scale=0.5]{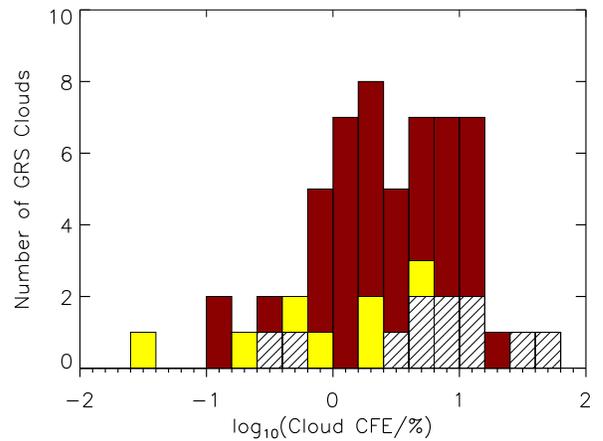}
\caption{Distribution of the clump formation efficiencies for individual GRS clouds with the tangent, background and foreground components depicted by the red, yellow and white hashed bars respectively.}
\label{CFEhist}
\end{center}
\end{figure}

Fig.~\ref{CFEhist} displays the distribution of CFEs of individual GRS clouds belonging to these three regions. It can be seen that, in addition to the cloud associated with the W43 H$\,\textsc{II}$ region, which is part of the tangent sample, there are also extreme-CFE clouds in the foreground and background samples. Since these are not affected by the presence of the Galactic Bar, the latter cannot be securely identified as the cause of the high CFE or SFE in W43. In fact a K--S test of the three samples shows no evidence that they are distributed differently, with a 20 per cent probability of being drawn from the same underlying population.

Additionally, an Anderson--Darling test of the combined samples shows no evidence that they are not distributed log-normally at the 1.5--$\sigma$ level. The implication of this is that, not only are clouds in the tangent region unlikely to be exceptional, despite the influence of the Galactic Bar, their star-forming properties are distributed randomly, so that the extreme-CFE clouds, including W43, are simply those in the wings of the log-normal distribution and are present only because the sample is large enough. This is not to say that they are not extreme, or potentially useful analogues for larger starburst systems, but it removes any need to argue for an exceptional environment associated with the Galactic Bar to explain their presence.

This conclusion is supported by Moore et al. (2012, in prep) who, by using the ratio of YSO luminosity to cloud mass as a measure of star-formation efficiency on kpc scales, found no large enhancement associated with the Scutum--Centaurus arm tangent, where W43 is found, despite strong peaks in both star-formation rate and molecular mass surface density in that region.

Observations of other barred spiral galaxies have found large amounts of molecular material and extreme star-forming regions at the ends of galactic bars (e.g. \citealt{b62}; \citealt{b51}; \citealt*{b58}) where the gas and stars in normal circular orbits are in co-rotation with the pattern speed of the bar (\citealt{b60}; \citealt{b61}). The gas in the bar, which is on long orbits up and down the length of the bar, is expected to interact with the circularly orbiting clouds and the gas at the leading edge of the bar is shocked \citep*{b59}. Bar ends might therefore be expected to produce unusual numbers of extreme star-forming regions from increased rates of cloud-cloud collisions which have been shown to induce star formation (\citealt{b65}; \citealt{b64}).

However, the idea that local enhancements in CFE and SFE may have very little to do with large-scale structures is consistent with the work of \citet{b57} who found no significant increase in SFE in spiral arms, relative to the inter-arm gas, in two Grand Design spiral galaxies. Since the CFE variations begin to disappear on scales significantly larger than the scale size of molecular clouds, the dominant mechanism affecting the SFE is more likely to be on the scale of individual clouds.

\section{Conclusions}

We have derived distances to 793 of the 806 BGPS sources within a 3 $\times$ 2 degree field centred at $\emph{l}$ = 30$\degr$, using a variety of techniques including  positional association with GRS molecular clouds with known distances and new kinematic distances using $^{13}$CO $J$ = 3 $\rightarrow$ 2 data, with H$\,\textsc{I}$ self-absorption and scale-height cutoffs to resolve ambiguities. The derived distances were used to calculate the masses for these 793 BGPS sources.

By separating the region into three distance-defined groups associated with separate Galactic structure features in the $\emph{l}$ = 30$\degr$ field, we are able to test how the major star formation metrics, such as the clump mass function (CMF) and clump formation efficiency (CFE), vary with proximity to features of large-scale Galactic structure. These three regions are the foreground spur/Sagittarius arm; the background Perseus and Sagittarius arms and the tangent of the Scutum--Centaurus arm and the Galactic Long Bar.

The clump mass functions of the three regions were found to be well fitted by a single power-law for clump masses greater than the completeness limits. The slopes were found to be -1.87 $\pm$ 0.13, -1.53 $\pm$ 0.10 and -1.48 $\pm$ 0.04 for the foreground, background and tangent regions respectively.  The slopes for the background and tangent regions are consistent with each other, however the foreground region is found to depart at the 3--$\sigma$ level. This is potential evidence for a steeper slope in the foreground component.

The clump formation efficiencies were also investigated with relation to their Galactic environments. The CFEs, defined as the ratio of clump masses to cloud masses, were found to be 13.00 $\pm$ 1.60, 8.52 $\pm$ 0.73 and 8.38 $\pm$ 0.30 per cent for the foreground, background and tangent regions respectively. As with the CMFs, the background and tangent regions are consistent with each other. However, the foreground CFE is found to be significantly higher at the 3--$\sigma$ level. This is surprising as we would have expected the CFE within the tangent region to be higher due to its position at the end of the Galactic Bar.

The median CFEs for the individual clouds were found to be 7.15 $\pm$ 4.95, 3.78 $\pm$ 3.32 and 3.27 $\pm$ 2.28 per cent for the foreground, background and tangent clouds respectively, with the mean values found to be 12.31 $\pm$ 4.86, 5.95 $\pm$ 2.26 and 5.38 $\pm$ 1.22 per cent. These are consistent within the uncertainties at an approximate 1--$\sigma$ level, implying that the clouds within the tangent region are not, on average, in a significantly different physical state to those in the other two galactic environments.

We also investigate changes in the CFE across the tangent region, finding large variations with localised maxima coincident with well-known H$\,\textsc{II}$ regions. Although some of these peaks are quite extreme, we find no evidence that the distribution of CFE values in the tangent region is distinct from those in the other two spiral-arm features and the combined distribution is found to be consistent with being log-normal. There is therefore no reason to believe that star formation is abnormally efficient in the Scutum-tangent clouds or proceeding differently there due to the location at the end of the Galactic Long Bar. This result may also imply that local triggering due to feedback is the dominant mechanism affecting the CFE and, consequently, the star-formation efficiency.

\section*{Acknowledgments}

This publication makes use of molecular line data from the Boston University-FCRAO Galactic Ring Survey (GRS). The GRS is a joint project of Boston University and Five College Radio Astronomy Observatory, funded by the National Science Foundation under grants AST-9800334, AST-0098562, \& AST-0100793. The James Clerk Maxwell Telescope is operated by The Joint Astronomy Centre on behalf of the Science and Technology Facilities Council of the United Kingdom, the Netherlands Organisation for Scientific Research, and the National Research Council of Canada. The National Radio Astronomy Observatory is a facility of the National Science Foundation operated under cooperative agreement by Associated Universities, Inc. DJE wishes to acknowledge an STFC PhD studentship for this work. LKM and TJTM were funded by STFC grant ST/G001847/1. RP is funded through a grant from the Natural Sciences and Engineering Research Council of Canada. This research has made use of NASA's Astrophysics Data System.

\bibliographystyle{mn2e}
\bibliography{ref}

\label{lastpage}

\end{document}